\documentclass[12pt]{iopart}

\usepackage{iopams}
\usepackage[dvips]{graphicx}

\begin{document}

\title{Equivalence of volume and temperature fluctuations
in power-law ensembles}

\author{Grzegorz Wilk}
\address{The Andrzej So{\l}tan Institute for Nuclear Studies,
Ho\.{z}a 69, 00681, Warsaw, Poland} \ead{wilk@fuw.edu.pl}
\author{Zbigniew W\l odarczyk}
\address{Institute of Physics, Jan Kochanowski University,
\'Swi\c{e}tokrzyska 15, 25-406 Kielce, Poland}
\ead{zbigniew.wlodarczyk@ujk.kielce.pl}

\begin{abstract}
Relativistic particle production often requires the use of Tsallis
statistics to account for the apparently power-like behavior of
transverse momenta observed in the data even at a few GeV/c. In
such an approach this behavior is attributed to some specific
intrinsic fluctuations of the temperature $T$ in the hadronizing
system and is fully accounted by the nonextensivity parameter $q$.
On the other hand, it was recently shown that similar power-law
spectra can also be obtained by introducing some specific volume
fluctuations, apparently without invoking the introduction of
Tsallis statistics. We demonstrate that, in fact, when the total
energy is kept constant, these volume fluctuations are equivalent
to temperature fluctuations and can be derived from them. In
addition, we show that fluctuations leading to multiparticle
power-law Tsallis distributions introduce specific correlations
between the considered particles. We then propose a possible way
to distinguish the fluctuations in each event from those occurring
from event-to-event. This could have applications in the analysis
of high density events at LHC (and especially in ALICE).
\end{abstract}

\pacs{25.75.Ag, 24.60.Ky, 24.10.Pa, 05.90.+m}

\maketitle

\section{\label{sec:I}Introduction}

Statistical modelling represents a widely used standard tool to
analyze multiparticle production processes \cite{MG_rev}. However,
this approach does not account for the possible intrinsic
nonstatistical fluctuations in the hadronizing system. These
usually result in a characteristic power-like behavior of the
single particle spectra and in the broadening of the corresponding
multiplicity distributions. Such fluctuations are important as
they can signal a possible phase transition(s) taking place in an
hadronizing system \cite{PhTr}, so to include such features, one
should base this modelling on the Tsallis statistics
\cite{T,WW_epja,BPU_epja}, leading to a Tsallis distribution,
$h_q(E)$, which accounts for such situations by introducing in
addition to the temperature $T$ a new parameter, $q > 1$. This
parameter is, as shown in \cite{WW,BJ}, directly connected to
fluctuations of the temperature. For $q \rightarrow 1$ one
recovers the usual Boltzmann-Gibbs distribution, $f(E)$:
\begin{eqnarray}
h_q(E) &=& \exp_q \left(-\frac{E}{T}\right) = \frac{2 -
q}{T}\left[1 - (1-q)\frac{E}{T}\right]^{\frac{1}{1-q}} \label{eq:Tsallis}\\
&\stackrel{q \rightarrow 1}{\Longrightarrow}& f(E) =
\frac{1}{T}\exp \left(-\frac{E}{T}\right), \label{eq:example}\\ &&
q-1=\frac{Var(1/T)}{\langle 1/T\rangle^2} .\label{eq:qinterpr}
\end{eqnarray}
 The most recent applications of this approach come from the PHENIX
Collaboration at RHIC \cite{PHENIX} and from CMS Collaboration at
LHC \cite{LHC_CMS} (see also a recent compilation
\cite{qcompilation}). One must admit at this point, that this
approach is subjected to a rather hot debate of whether it is
consistent with equilibrium thermodynamics or else it is only a
handy way to a phenomenological description of some intrinsic
fluctuations in the system \cite{debate}. However, as was recently
demonstrated on general grounds in \cite{M}, fluctuation phenomena
can be incorporated into a traditional presentation of
thermodynamic and the Tsallis distribution (\ref{eq:example})
belongs to the class of general admissible distributions which
satisfy thermodynamic consistency conditions. They are therefore a
natural extension of the usual Boltzman-Gibbs canonical
distribution (\ref{eq:example})\footnote{The recent generalization
of classical thermodynamics to a nonextensive case presented in
\cite{BUS} should be noticed in this context as well. It is worth
mentioning that, in addition to applications mentioned above, the
nonextensive approach has also been applied to hydrodynamical
models \cite{Hydro} and to investigations of dense nuclear matter,
cf., for example, \cite{DNM}.}.

In fact, as was shown in \cite{WW}, assuming some simple diffusion
picture as responsible for temperature equalization in a
nonhomogeneous heat bath (in which local temperature, $T$,
fluctuates from point to point around some equilibrium value,
$T_0$) one gets the evolution of $T$ in the form of a Langevin
stochastic equation and distribution of $1/T$, $g(1/T)$, as a
solution of the corresponding Fokker-Planck equation. It turns out
that $g(1/T)$ has the form of a gamma distribution,
\begin{eqnarray}
g(1/T) = \frac{1}{\Gamma\left(\frac{1}{q - 1}\right)}\frac{T_0}{q
- 1}\left(\frac{1}{q - 1}\frac{T_0}{T}\right)^{\frac{2 - q}{q -
1}}\cdot \exp\left( - \frac{1}{q - 1}\frac{T_0}{T}\right).
\label{eq:gamma}
\end{eqnarray}
Convoluting $\exp (- E/T)$ with such $g(1/T)$ one immediately gets
Tsallis distribution (\ref{eq:Tsallis}) with a well physically
defined parameter $q$: according to Eq. (\ref{eq:qinterpr}) it is
entirely given by the temperature fluctuation pattern, which in
turn is fully described by the parameters entering this basic
diffusion process (like, for example, conductance and specific
heat of the hadronic matter consisting this nonhomogeneous heat
bath, cf., \cite{WW} for details). This approach was recently
generalized to account for the possibility of transferring energy
from/to a heat bath with a new parameter characterizing the
corresponding viscosity entering into the definition of $q$ (it
appears to be important for AA applications \cite{WW_epja,WWprc}
and for cosmic ray physics \cite{WWcosmic}; we shall not discuss
this issue here) \footnote{In \cite{GH} a similar suggestion of
the extension of standard concept of statistical ensembles was
proposed independently. A class of ensembles with extensive
(rather than intensive) quantities fluctuating according to an
externally given distribution was discussed. There is also a
purely phenomenological approach treating the occurrence of a
Tsallis distribution as a manifestation of the so called
superstatistics, see \cite{Ss}.}.

In the next Section, we shall discuss the correspondence between
fluctuations of volume $V$ proposed in \cite{Vfluct} and the
presented above fluctuations of temperature $T$ (both result in
power-like distributions). Section \ref{sec:III} is devoted to a
discussion of some specific $q$-induced correlations and
fluctuations. Section \ref{sec:IV} is a summary.

\section{\label{sec:II}Fluctuations of $V$ or $T$?}

We start by stressing that the form of $g(1/T)$ as given by Eq.
(\ref{eq:gamma}) is not assumed, but derived from the properties
of the underlying physical process in the nonhomogeneous heat
bath. Apparently, the same results in what concerns the power-like
character of single particle spectra and the broadening of the
corresponding multiplicity distributions, $P(N)$, were obtained in
\cite{Vfluct} without resorting to Tsallis statistics. It was
assumed there that the volume $V$ fluctuates in scale invariant
way following the KNO form of $P(N)$ deduced from the experiment
\cite{KNO}\footnote{It should be noticed that UA5 data \cite{UA5}
demonstrated that KNO scaling is broken via the energy dependence
of the parameter $k$. In fact, as shown in \cite{GG}, $k^{-1} =
-0.104 + 0.058\ln \sqrt{s}$ . Therefore, in the scenario with
fluctuations of the volume $V$, the scaling KNO form of the $P(N)$
used to model these fluctuations is a rather rough simplification.
On the contrary, in the scenario of the temperature $T$
fluctuations, $ P(N)$ is given by a Negative Binomial
Distribution, which adequately describes the data. }. We shall now
demonstrate that, for the case of constant energy $E$ considered
in \cite{Vfluct}, both approaches are equivalent in the sense that
one can start from fluctuations of $V$ and recover fluctuations of
$T$ as discussed above, or else, one can start from fluctuations
of $T$ as given by $g(1/T)$ and recover fluctuations of $V$ as
assumed in \cite{Vfluct}.

Following the approach of \cite{Vfluct}, for constant total
energy, $E=const$, when both the volume $V$ and temperature $T$
are related via $E \sim VT^4$, i.e., when
\begin{equation}
T = \langle T\rangle\left( \frac{\langle V \rangle}{V}
\right)^{\frac{1}{4}}, \label{eq:TV}
\end{equation}
the mean multiplicity in the microcanonical  ensemble (MCE),
$\bar{N}$, can be written as
\begin{equation}
\bar{N} = \langle N\rangle\cdot \frac{V}{\langle V\rangle}\left(
\frac{T}{\langle T\rangle}\right)^3 = \langle N\rangle\ y\quad
{\rm where}\quad  y = \left( \frac{V}{\langle
V\rangle}\right)^{1/4}. \label{eq:y} \label{eq:barNV}
\end{equation}
This relation points to the KNO scaling form of the multiplicity
distribution as good candidate for distribution of $y$, which was
therefore assumed to be given by \cite{KNO}
\begin{equation}
\psi(y) = \frac{k^k}{\Gamma(k)}y^{k-1}\exp( - ky).
\label{eq:distry}
\end{equation}
From Eqs. (\ref{eq:barNV}) and (\ref{eq:TV}) one gets that
\begin{equation}
y = \frac{\langle T\rangle}{T}. \label{eq:qT}
\end{equation}
This means that $T$ fluctuates as well according to the
distribution $\psi(\langle T\rangle/T)$. The power-like form of
the single particle spectra then follows immediately, all
apparently without invoking any reference to Tsallis statistics.
This completes the proposed picture which now comprises both
fluctuations presented in the multiplicity distribution $P(N)$
(from the scaling form of which one deduces the shape of volume
fluctuations) and the power-like behavior of single particle
spectra emerging because of temperature fluctuations that follow.
Notice now that that $\psi(\langle T\rangle/T)$ assumed here is,
in fact, the same distribution as $g(1/T)$ derived in Eq.
(\ref{eq:gamma}) (with $k = 1/(q - 1)$, see also Eq.
(\ref{eq:NBD}) below). So we obtain from $V$ fluctuations the $T$
fluctuations with the same functional form but now without the
physical background behind Eq. (\ref{eq:gamma}) mentioned above.

However, we can proceed in reverse order and obtain from $T$
fluctuations (\ref{eq:gamma}) introduced in Section \ref{sec:I}
the fluctuations of $V$ introduced in \cite{Vfluct}, including the
broadening of the corresponding $P(N)$ which takes the form of a
NB distribution. This point has been already shown in \cite{fluct}
and we shall quote here its main points for the sake of
completeness.

As was proved there, $T$ fluctuations in the form of Eq.
(\ref{eq:gamma}), discussed in Section \ref{sec:I}, result in a
specific broadening of the corresponding multiplicity
distributions, $P(N)$, which evolve from the poissonian form
characteristic for exponential distributions to the negative
binomial (NB) form observed for Tsallis distributions. One starts
from the known fact that whenever we have $N$ independently
produced secondaries with energies $\{ E_{i=1,\dots,N}\}$ taken
from the exponential distribution $f(E)$, cf. Eq.
(\ref{eq:example}), i.e., when the corresponding joint
distribution is given by
\begin{equation}
f\left( \{ E_{i=1,\dots,N}\}\right) =
\frac{1}{\lambda^N}\exp\left( - \frac{1}{\lambda}\sum^{N}_{i=1}
E_i\right), \label{eq:PNBG}
\end{equation}
and whenever
\begin{equation}
\sum^N_{i=0} E_i \leq E \leq \sum^{N+1}_{i=0} E_i,
\label{eq:condition}
\end{equation}
the corresponding multiplicity distribution is poissonian,
\begin{equation}
P(N) = \frac{\left( \bar{N}\right)^N}{N!} \exp\left ( -
\bar{N}\right) \quad {\rm where}\quad \bar{N} =\frac{E}{\lambda}.
\label{eq:Poisson}
\end{equation}
On the other hand, whenever in some process $N$ particles with
energies  $\{ E_{i=1,\dots,N}\}$ are distributed according to the
joint $N$-particle Tsallis distribution,
\begin{equation}
h\left(\{ E_{i=1,\dots,N}\} \right)\! =\! C_N\left[ 1-
(1-q)\frac{\sum^N_{i=1} E_i }{\lambda} \right]^{\frac{1}{1-q}+1-N}
\label{eq:NTsallis}
\end{equation}
(for which the corresponding one particle Tsallis distribution
function in Eq. (\ref{eq:Tsallis}) is marginal distribution),
then, under the same condition (\ref{eq:condition}), the
corresponding multiplicity distribution is the NB distribution,
\begin{equation}
P(N)\, =\, \frac{\Gamma(N+k)}{\Gamma(N+1)\Gamma(k)}\frac{\left(
\frac{\langle N\rangle}{k}\right)^N}{\left( 1 + \frac{\langle
N\rangle}{k}\right)^{(N+k)}};\quad {\rm where}\quad
k=\frac{1}{q-1}.\label{eq:NBD}
\end{equation}
Notice that, in the limiting case of $q\rightarrow 1$, one has
$k\rightarrow \infty$ and (\ref{eq:NBD}) becomes a poissonian
distribution (\ref{eq:Poisson}), whereas for $q\rightarrow 2$ on
has $k\rightarrow 1$ and (\ref{eq:NBD}) becomes a geometrical
distribution. It is easy to show that for large values of $N$ and
$\langle N\rangle$ one obtains from Eq. (\ref{eq:NBD}) its scaling
form,
\begin{equation}
\langle N\rangle P(N) \cong  \psi\left( z=\frac{N}{\langle
N\rangle} \right) = \frac{k^k}{\Gamma(k)} z^{k-1}\exp( - kz),
\label{eq:scalingform}
\end{equation}
in which one recognizes a particular expression of
Koba-Nielsen-Olesen (KNO) scaling \cite{KNO} assumed in
\cite{Vfluct} to also describe the volume fluctuations, cf. Eq.
(\ref{eq:distry}). This result closes the demonstration that,
under the condition of constancy of total energy used here, $T$
and $V$ fluctuations are equivalent \footnote{In fact one can
argue that the scaling form of $P(N)$ visible in experiments
points to the necessity of describing multiparticle production
processes by means of Tsallis statistics. It is worth mentioning
at this point that the connection between $q$ and $k$ was first
discovered in \cite{RWW} when fitting $p\bar{p}$ data for
different energies by means of the Tsallis formula
(\ref{eq:Tsallis}). The resulting energy dependence of parameter
$q$ turned out to coincide with that of $1/k$ of the respective
NBD fits to corresponding $P(N)$. It was then realized that
fluctuations of $\bar{N}$ in the poissonian distribution
(\ref{eq:Poisson}) taken in the form of $\psi(\bar{N}/<N>)$, Eq.
(\ref{eq:scalingform}), lead to the NB distribution
(\ref{eq:NBD}).}.

We close this Section with the following remarks. As was said
above, $T$ fluctuations are derived from the more realistic
description of the nonhomogeneous heat bath, and therefore
parameter $q$ and the Tsallis distribution, Eq.
(\ref{eq:Tsallis}), reflect the physics of this heat bath.
However, one can argue that, on this deeper level, this physics is
nothing more than some phenomenological modelling, assumptions of
which are reflected in $q$. The $V$ fluctuations approach uses
instead as its input the experimental knowledge of the scaling
properties of particle multiplicity distributions, $P(N)$,
assuming that fluctuations presented there are transmitted to
fluctuations of the volume. In fact, fluctuations of $V$ could
have some deeper phenomenological foundation, not mentioned in
\cite{Vfluct}. Namely, it is known that one observes
experimentally a variation of the emitting radius (evaluated from
the Bose-Einstein correlation analysis) with the charged
multiplicity of the event, see, for example, \cite{R_fluct}. An
increase of about $10$ \% of the radius when the multiplicity
increases from $10$ to $40$ charged hadrons in the final state was
reported. Unfortunately, the quality of data does not allow us to
precisely determine the power index of the volume dependence. It
is also remarkable that both the energy density, $\rho_E=E/V$, and
particle density, $\rho_N = N/V$, decrease for large multiplicity
events. For $N/\langle N\rangle \sim y$ one observes
$\rho_E/\langle \rho_E\rangle \sim y^{-4}$ and $\rho_N/\langle
\rho_N\rangle \sim y^{-3}$. All these features deserve further
consideration and should be checked in LHC experiments, especially
in ALICE, which is dedicated to heavy ion collision.

\section{\label{sec:III}Some consequences of $q$ statistics}

We would like to close with a short discussion of some
consequences of $q$-statistics which can {\it a priori} be
subjected to experimental verification: the $q$-induced
correlations and event-by-event fluctuations.

\begin{figure}[h]
\includegraphics[width=8cm]{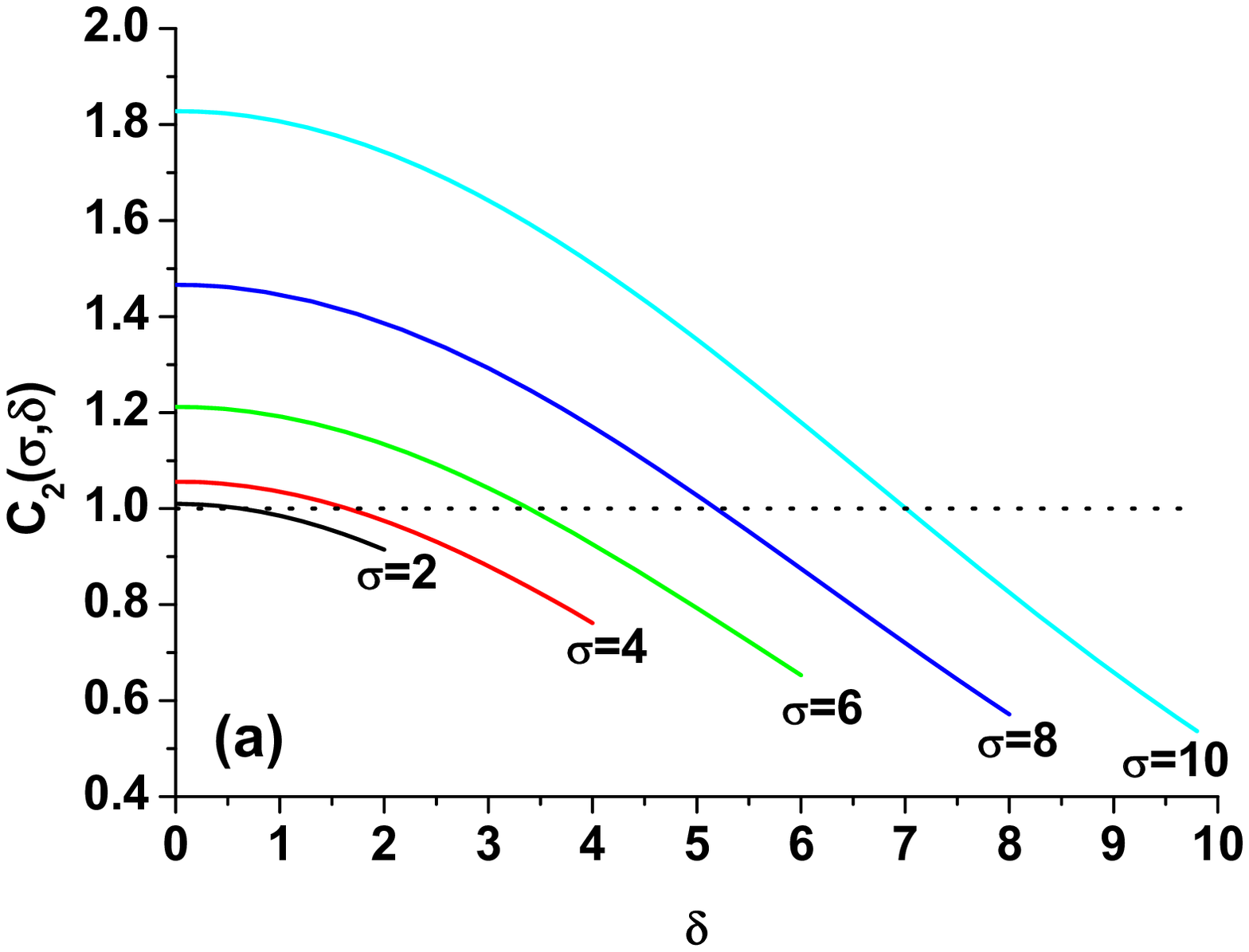}
\includegraphics[width=8cm]{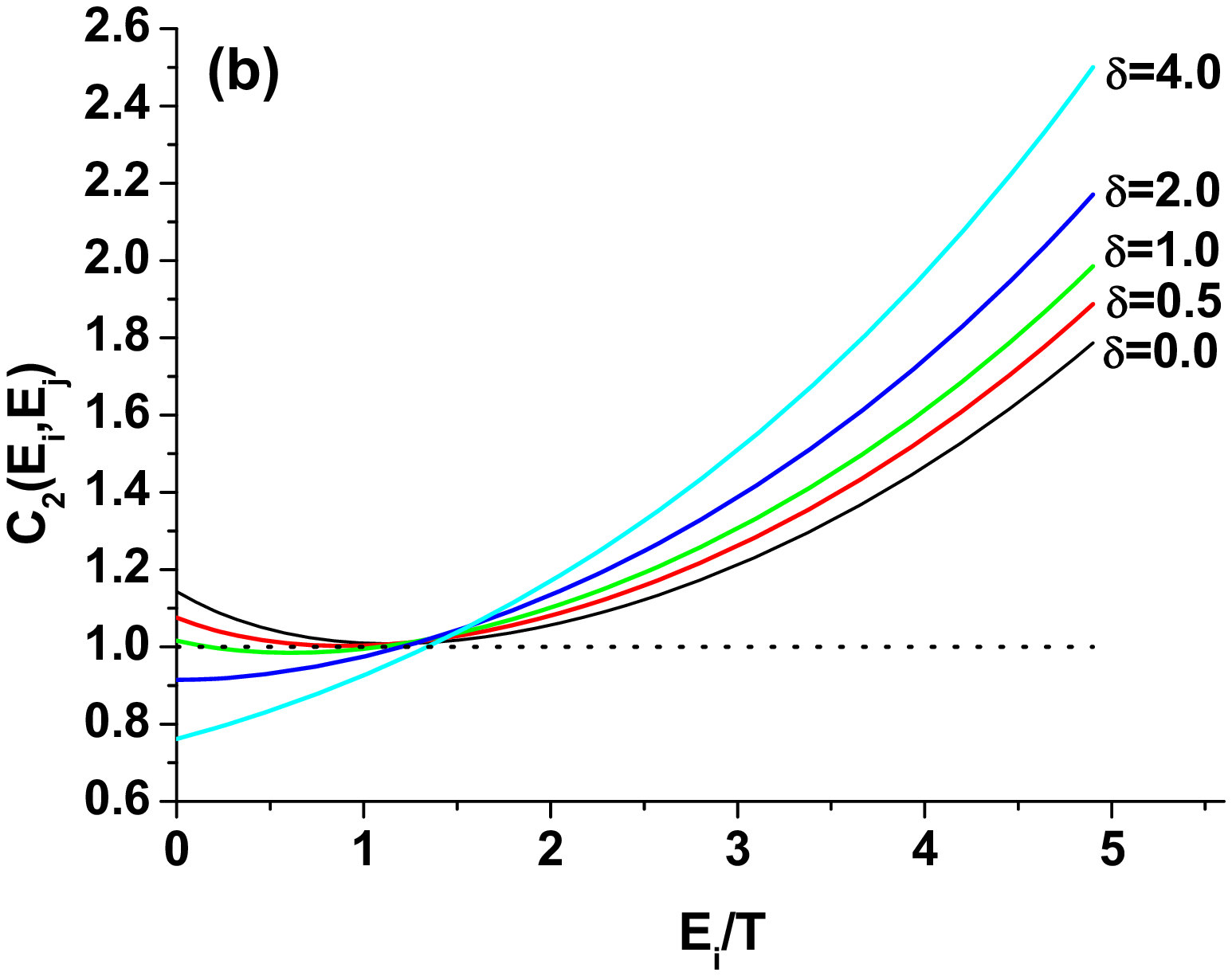}
 \caption{(Color online). Two-particle correlation function
           $C_2\left( E_i, E_j\right)$
           for Tsallis distribution with the parameter $q=1.125$.
           }
 \label{fig1}
\end{figure}

The $q$-induced correlations occur in a natural way in an
$N$-particle Tsallis distribution introduced in Eq.
(\ref{eq:NTsallis}). For Boltzmann-Gibbs statistics, for $N$
independently produced particles, the joint distribution
(\ref{eq:PNBG}) can be written in factorizable form as a simple
product of single particle distributions,
\begin{equation}
f\left( \{E_{i=1,\dots,N}\}\right) = \prod_{i=1}^N f\left(
E_i\right). \label{eq:NBG}
\end{equation}
However, such a product of single particle Tsallis distributions
does not result in an $N$-particle Tsallis distribution
\cite{fluct}. To get Eq. (\ref{eq:NTsallis}) one has to fluctuate
the temperature in the distribution $f\left(
\{E_{i=1,\dots,N}\}\right)$ above, i.e.,
\begin{equation}
h\left( \{E_{i=1,\dots,N}\}\right)\!\! =\!\! \int^{\infty}_0\!\!\!
f\left( \{E_{i=1,\dots,N}\}\right)g(1/T)d(1/T).
\end{equation}
This procedure introduces correlations between particles. The
corresponding covariance, $Cov\left( E_i,E_j \right)$, and
correlation coefficient, $\rho$, for energies are equal to
\begin{eqnarray}
&& Cov\left( E_i,E_j \right) =
\frac{\lambda^2(q-1)}{(3-2q)^2(4-3q)}, \label{eq:covariance}\\
&& \rho = \frac{1}{2-q}-1.\label{eq:rho}
\end{eqnarray}
As an illustrative example, we calculate the two-particle
correlation function
\begin{equation}
C_2\left( E_i,E_j \right) = \frac{h\left( E_i,E_j\right)}{h\left(
E_i\right)h\left( E_j\right)}. \label{eq:C2}
\end{equation}
This is shown in Fig. \ref{fig1} for different values of variables
$\delta$ and $\sigma$,
\begin{equation}
 \delta = \frac{ E_i - E_j }{T}\quad {\rm and}\quad \sigma = \frac{ E_i
+ E_j}{T}. \label{eq:def}
\end{equation}
Notwithstanding the rather complicated dependencies shown in Fig.
\ref{fig1}, the distance distribution, $h_q$ defined as
\begin{equation}
h_q (\delta) = \int^{\infty}_0 h_q\left( E_i,E_i+\delta\right)
dE_i \label{eq:distance}
\end{equation}
for variables sampled from the joint distribution
(\ref{eq:NTsallis}) is given by a Tsallis distribution
\begin{equation}
h_q (\delta) = \frac{2-q}{2}[1 - (1-q)\delta]^{\frac{1}{1-q}}
\label{eq:distance1}
\end{equation}
(analogously as the distance distribution for the exponentially
distributed variables is given by exponential distribution, to
which (\ref{eq:distance}) converges for $q\rightarrow 1$).

In what concerns event-by-event fluctuations there are two types
of fluctuations one can encounter, namely {\it fluctuations from
event to event} or {\it fluctuations in each event}.  Two
scenarios are possible here.
\begin{itemize}
\item[$(i)$] In the first, $T$ (and/or $V$) are constant in each
event. However, because of different initial conditions, they
fluctuate from event to event. In this case, in each event one
should find an exponential dependence (\ref{eq:example}) with
$T=T_k$ and a possible departure from it could occur only after
averaging over all $N_{ev}$ events, $k=1,2,\dots,N_{ev}$. It will
reflect fluctuations originating in the different initial
conditions collision from which a given event originates. Only
inclusive distributions will be described by a Tsallis formula
(\ref{eq:Tsallis}). Such a situation was discussed above.
\item[$(ii)$] In the other scenario, $T$ fluctuates in each event
around some value $T_0$. In this case one should observe a
departure from the exponential behavior already on the single
event level and it should follow Eq. (\ref{eq:Tsallis}) with $q >
1$. This reflects the situation when, due to some intrinsically
dynamical reasons, different parts of a given event can have
different temperatures \cite{WW,WW_epja}. For volume fluctuations
such a scenario seems to be unrealistic.
\end{itemize}

In \cite{ebefluct} we argued that event-by-event analysis of
multiparticle production data are an ideal place to search for a
possible fluctuation of the temperature characterizing a
hadronizing source using the thermodynamical approach. Namely, an
analysis of the transverse momentum spectra in $Pb+Pb$ collisions
at TeV energies of LHC (mainly in ALICE experiment) should allow
us to distinguish between both scenarios listed above. The LHC is
designed for colliding proton-proton and nucleus-nucleus beams up
to $\sqrt{s} = 14$ TeV \cite{ALICE}. Collisions at these
unprecedented high energies will provide opportunities for new
types of analysis. In proton-proton collisions at the highest
possible energy, the expected charged particle multiplicity is
only $\sim 10$ at the midrapidity region ( $| \eta| < 0.5$) and it
is roughly five times bigger in the full rapidity region
\cite{full} (the ALICE experiment has the possibility of measuring
the distributions over the $-5.0 < \eta < 3.5$ range, but the CMS
and ATLAS experiments have a more limited coverage of $|\eta| <
2.5$ units). The most important fact is that, in heavy ion
collisions we have $\sim A$ higher multiplicities (for central
$Pb+Pb$ collisions one expects $\sim 2500$ particles at allowed
rapidity acceptance region). This is enough to analyze
event-by-event distributions over $3$ orders of magnitude, which
should allow us to determine the shape of the distribution in a
single event. Moreover,  in such circumstances we should be able
to also construct the distribution $dN/d\delta$ for $\delta =
p_{Ti} - p_{Tj}$ for $N(N-1) \cong 6\cdot10^6$ pairs in an event,
over $6$ orders of magnitude and using Eq. (\ref{eq:distance})
test the above possible scenarios. Eq. (\ref{eq:distance}) tells
us that, instead of distributions of $p_T$, $dN/dp_T$, one can use
distributions $dN/d\delta$ and look whether it follows the Tsallis
form on an event-by-event basis. Because for Eq.
(\ref{eq:distance}) one has $N(N-1)$ entries to be compared with
only $N$ for $p_T$ distributions, one expects that the
distribution Eq. (\ref{eq:distance}) will reach further and it
would be easier to differentiate between the Tsallis distribution
and the usual exponential one.

\begin{figure}[h]
\includegraphics[width=8cm]{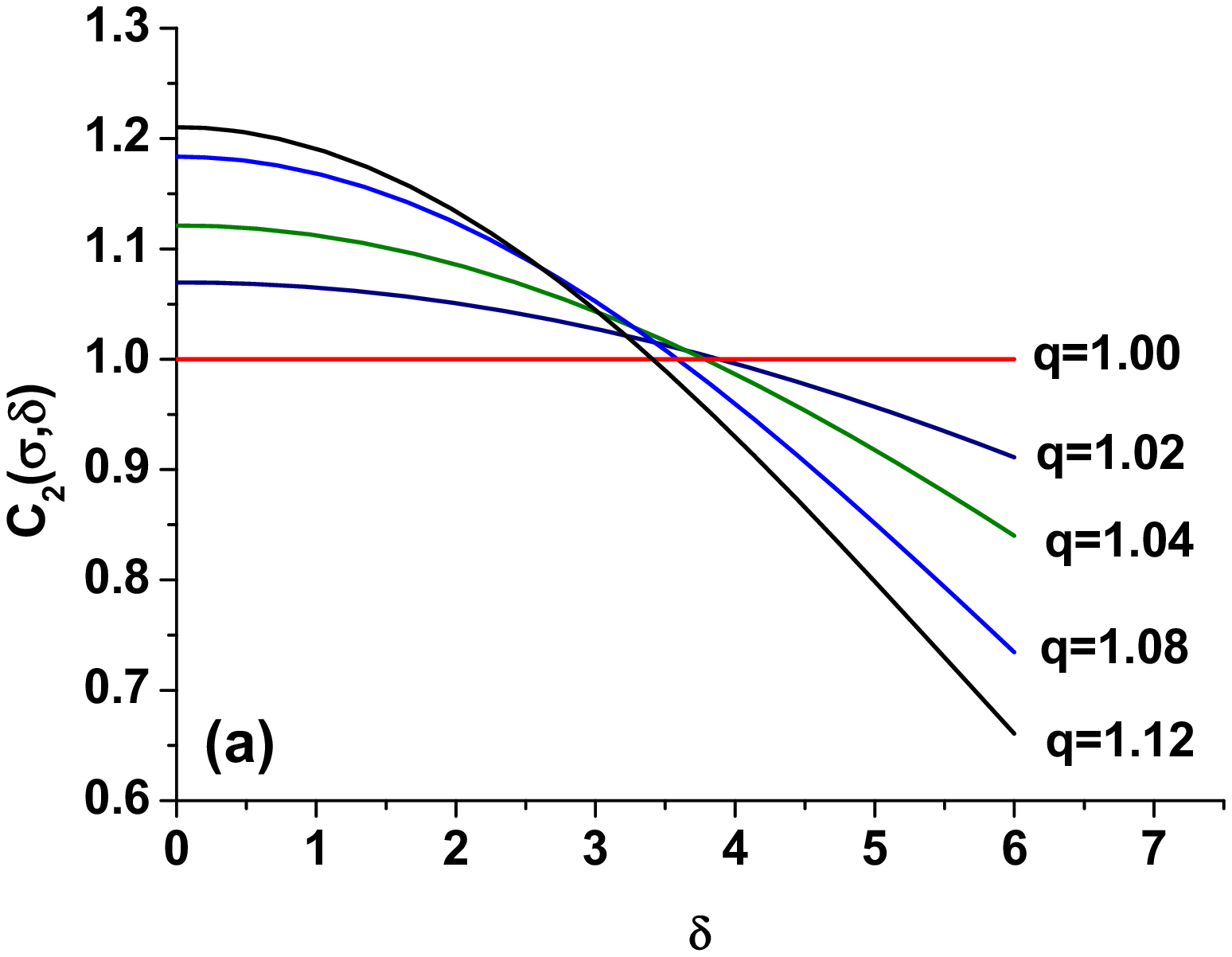}
\includegraphics[width=8cm]{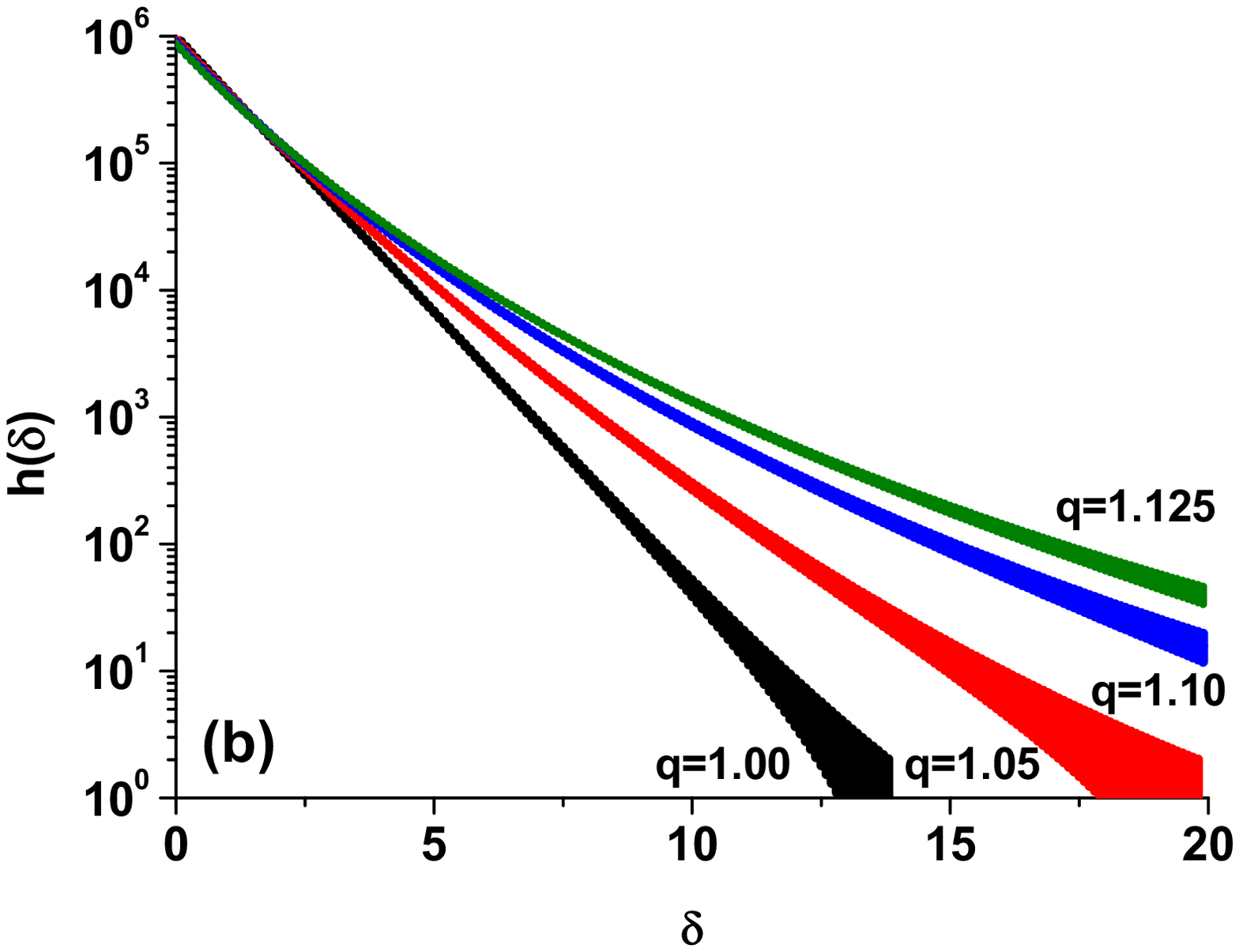}
 \caption{(Color online). The sensitivity to the parameter $q$ of
 $(a)$ the correlation function $C_2(\sigma, \delta)$ with
 $\sigma = 6$ and $(b)$ the distribution of $\delta$, $h(\delta)$,
 with error band corresponding to the event multiplicity $N=10^3$.
           }
 \label{fig2}
\end{figure}

Fig. \ref{fig2}a presents the sensitivity of the correlation
function $C_2(\sigma,\delta)$ to different values of the parameter
$q$ chosen. Similarly, Fig. \ref{fig2}b presents sensitivity of
the proposed method in the event-by-event analysis on the example
of distribution $h(\delta)$ plotted for different values of the
parameter $q$ with error bands corresponding to the event with
multiplicity $N = 10^3$.

We close this Section with a few remarks on what can be deduced
from existing experimental data. To separate the two scenarios
presented here one could, for example, use the analysis of
event-by-event fluctuations of the average transverse momenta
$\bar{p_T}$ presented at RHIC experiments. STAR data \cite{STAR}
show that $Var\left(\bar{p_T}\right)/\langle \bar{p_T}\rangle^2
\cong 10^{-4}$, where $\bar{p_T}$ denotes averaging in an event
and $\langle \bar{p_T}\rangle$ is the overall event average. In
the above estimation $Var\left(\bar{p_T}\right) =
Var\left(\bar{p_T}\right)_{data} -
Var\left(\bar{p_T}\right)_{mixed}$. STAR \cite{STAR} measured also
the quantity $ \langle \Delta p_{Ti}\Delta p_{Tj}\rangle =
Cov\left(p_{Ti},p_{Tj}\right) - Var\left(\bar{p_T}\right)$. In
this case  $\Delta p_{Ti} = p_{Ti} - \langle \bar{p_T}\rangle$ was
estimated without using mixed event and the resulting value of
fluctuations was the same as above (cf. also \cite{fluct}).
Therefore, in the scenario of volume fluctuations (or temperature
fluctuations occurring from event to event), assuming that
$\bar{p_T} \propto T$, we obtain $Var(T)/\langle T\rangle^2 =
Var\left( \bar{p_T}\right)/\langle \bar{p_T} \rangle^2 \cong
10^{-4}$, leading to $q - 1 \cong 10^{-4}$. This is much smaller
than the value $q - 1 \cong 10^{-2}$ estimated from transverse
momenta distributions \cite{WW_epja,PHENIX,LHC_CMS}. However, if
the temperature $T$ fluctuates within the event (for example, is
different in every inter-nuclear collisions), we expect that for
the $N_P$ projectile participants $Var\left( \bar{p_T}\right)
\cong Var(T)/N_P$. In this case, for $N_P\cong 100$ the relative
fluctuations are $Var(T)/\langle T\rangle^2 \cong 10^{-2}$, i.e.,
are comparable to those obtained from the transverse momentum
distributions mentioned above. Similar estimations could be made
also for \cite{PHENIX1} data. A more detailed analysis of this
type is outside the scope of this paper and will be presented
elsewhere.

\section{\label{sec:IV}Summary}

To summarize: we have demonstrated that two approaches to
fluctuation phenomena observed in multiparticle production
processes as power law distributions or broadening of the
corresponding multiplicity distributions, the one based on
temperature fluctuations \cite{WW,fluct} and the one based on
volume fluctuations \cite{Vfluct} can be regarded, as long as the
corresponding total energy is kept constant, to be equivalent. One
can be deduced from the other. We have also shown that
fluctuations which lead to multiparticle power-law distributions
introduce some specific correlations between particles in the
ensemble of particles considered and propose a way to distinguish
the fluctuations in each event from those occurring from
event-to-event by analyzing the distance distribution, Eq.
(\ref{eq:distance}). On the other hand, it seems that the already
existing RHIC data on $p_T$ correlations \cite{STAR,PHENIX1} can
also be useful to obtain such information (this point demands,
however, a more detailed study).

\section*{Acknowledgment}

Partial support (GW) of the Ministry of Science and Higher
Education under contract DPN/N97/CERN/2009 is acknowledged.

\end{document}